\documentclass[fleqn,twoside]{article}
\usepackage{amsfonts,bezier,epsf}

\newcommand{\bls}[1]{\renewcommand{\baselinestretch}{#1}}
\def\noi{\noindent}

\makeatletter
\renewcommand{\section}{\@startsection{section}{1}{0pt}%
        {-3.5ex plus -1ex minus -.2ex}{2.3ex plus .2ex}%
        {\large\bf\protect\raggedright}}

\renewcommand{\subsection}{\@startsection{subsection}{2}{0pt}%
        {-3ex plus -1ex minus -.2ex}{1.4ex plus .2ex}%
        {\normalsize\bf\protect\raggedright}}

\renewcommand{\@oddhead}{\raisebox{0pt}[\headheight][0pt]{%
   \vbox{\hbox to\textwidth{\rightmark \hfil \rm \thepage \strut}\hrule}}}
\renewcommand{\@evenhead}{\raisebox{0pt}[\headheight][0pt]{%
   \vbox{\hbox to\textwidth{\thepage \hfil \leftmark \strut}\hrule}}}

\newcommand{\heads}[2]{\markboth{\protect\small\it #1}{\protect\small\it #2}}
\makeatother

\newcommand{\Title}[1]{\noi {\Large #1} \\}
\newcommand{\Author}[2]{\noi{\large\bf #1}\\[2ex]\noindent{\it #2}\\}
\newcommand{\Abstract}[1]{\vskip 2mm \begin{center}
        \parbox{16.4cm}{\small\noi #1} \end{center}\medskip}
\newcommand{\foom}[1]{\protect\footnotemark[#1]}

\newcommand{\email}[2]{\footnotetext[#1]{e-mail: #2}
	\addtocounter{footnote}{1}}

\newcommand{\Ref}[1]{Ref.\,\cite{#1}}

\newcommand{\sect}[1]{Sec.\,#1}

\def\nq{\hspace*{-1em}}
\def\nqq{\hspace*{-2em}}
\def\nhq{\hspace*{-0.5em}}

\def\cm{\hspace*{1cm}}
\def\inch{\hspace*{1in}}

\newcommand{\Theorem}[2]{\medskip\noi {\bf #1. \ }{\it #2}\medskip}
\def\Description#1{\begin{description}\itemsep -2pt
	#1 \end{description}\vspace{-5pt}}

\newcommand{\Picture}[3]{
	\begin{figure} 	\centering \unitlength=1mm
	\begin{picture}(84,#1)
		\put(0,0){\line(0,1){#1}}            
		\put(0,0){\line(1,0){84}}
		\put(84,0){\line(0,1){#1}}
		\put(0,#1){\line(1,0){84}}
	\put(0,0){#2}                       \end{picture}
        \caption{\protect\small #3}  \smallskip \hrule \end{figure}
	}

\def\eq{Eq.\,}
\def\eqs{Eqs.\,}
\def\beq{\begin{equation}}
\def\eeq{\end{equation}}
\def\bear{\begin{eqnarray}}
\def\al{&\nhq}
\def\lal{&&\nqq {}}               
\def\bearr{\bear \lal}
\def\ear{\end{eqnarray}}

\def\dst{\displaystyle}
\def\tst{\textstyle}

\def\nn{\nonumber\\ {}}

\def\nnn{\nonumber\\ \lal }
\def\nnnv{\nonumber\\[5pt] \lal }
\def\yy{\\[5pt] {}}

\def\eql{\al =\al}

\def\eqdef{\stackrel{\rm def}=}
\def\e{{\,\rm e}}
\def\d{\partial}

\def\sign{\mathop{\rm sign}\nolimits}

\def\const{{\rm const}}
\def\Half{{\dst\frac{1}{2}}}
\def\half{{\tst\frac{1}{2}}}

\def\DAL{\mathop{\raisebox{3pt}{\large\fbox{}}}\nolimits}

\def\Jl#1#2{{\it #1\/} {\bf #2},\ }

\def\CQG#1 {\Jl{Class. Qu. Grav.}{#1}}
\def\DAN#1 {\Jl{Dokl. AN SSSR}{#1}}
\def\GC#1 {\Jl{Grav. \& Cosmol.}{#1}}
\def\GRG#1 {\Jl{Gen. Rel. Grav.}{#1}}
\def\JETF#1 {\Jl{Zh. Eksp. Teor. Fiz.}{#1}}
\def\JMP#1 {\Jl{J. Math. Phys.}{#1}}
\def\NPB#1 {\Jl{Nucl. Phys.}{B\ #1}}
\def\PLA#1 {\Jl{Phys. Lett.}{#1A}}
\def\PLB#1 {\Jl{Phys. Lett.}{#1B}}
\def\PRD#1 {\Jl{Phys. Rev.}{D\ #1}}
\def\PRL#1 {\Jl{Phys. Rev. Lett.}{#1}}

\def\GR{general relativity}

\def\ssph{static, spherically symmetric}
\def\fig{Fig.\,}
\def\bh{black hole}
\def\bhs{black holes}

\def\Sch{Schwarzschild}
\def\dS{de Sitter}
\def\adS{anti-de Sitter}

\def\mn{_{\mu\nu}}
\def\MN{^{\mu\nu}}
\def\mN{_\mu^\nu}

\def\od{{\overline d}}
\def\og{{\overline g}}

\def\cA{{\cal A}}
\def\cR{{\cal R}}
\def\ocR{\overline{\cal R}}

\def\M{{\mathbb M}}

\def\R{{\mathbb R}}
\def\S{{\mathbb S}}
\def\Z{{\mathbb Z}}

\def\Df{\Delta\phi}

\def\ME {\mbox{$\M_{\rm E}$}}
\def\MJ {\mbox{$\M_{\rm J}$}}
\def\oM{{\overline \M}}
\def\map{\ \longleftrightarrow\ }
\def\Str{\mbox{$\S_{\rm trans}$}}

\heads{K.A. Bronnikov}
{Scalar-tensor gravity and conformal continuations}

\bls{0.98}

\begin{document}
\twocolumn[
\thispagestyle{empty}
	\begin{flushright}\unitlength=1pt
	\begin{picture}(130,25)
   		\put(0,40){\large\bf gr-qc/0204001}
	\end{picture}
	\end{flushright}

\vspace*{-12mm}
\Title
{\bf Scalar-tensor gravity and conformal continuations}

\Author{K.A. Bronnikov\foom 1}
{Centre for Gravitation and Fundamental. Metrology, VNIIMS,
        3-1 M. Ulyanovoy St., Moscow 117313, Russia;\\
Institute of Gravitation and Cosmology, PFUR,
        6 Miklukho-Maklaya St., Moscow 117198, Russia}

\Abstract
    {Global properties of vacuum static, spherically symmetric
     configurations are studied in a general class of scalar-tensor theories
     (STT) of gravity in various dimensions. The conformal mapping between
     the Jordan and Einstein frames is used as a tool. Necessary and
     sufficient conditions are found for the existence of solutions
     admitting a conformal continuation (CC). The latter means that a
     singularity in the Einstein-frame manifold maps to a regular surface
     \Str\ in the Jordan frame, and the solution is then continued beyond
     this surface. \Str\ can be an ordinary regular sphere or a horizon; it
     is found that in the second case \Str\ connects two epochs of a
     Kantowski-Sachs type cosmology. It is shown that the list of possible
     types of global causal structure of vacuum space-times in any STT, with
     any potential function $U(\phi)$, is the same as in \GR\ with a
     cosmological constant. This is even true for conformally continued
     solutions. A traversable wormhole is shown to be one of the generic
     structures created as a result of CC. Two explicit examples are
     presented: the known solution for a conformal field in \GR,
     illustrating the emergence of singularities and wormholes due to CC,
     and a nonsingular 3-dimensional model with an infinite sequence of CCs.
     }

] 
\email 1 {kb@rgs.mccme.ru}

\section {Introduction}    

    Scalar fields with various potentials are of great significance
    in various branches of theoretical physics and cosmology: it is
    sufficient to mention, e.g., the Higgs field in particle theory and
    numerous quintessence models in modern cosmology. It is thus highly
    desirable to know which kinds of gravitationally self-bound
    configurations can be formed by such fields.

    This paper continues the study of global properties of \ssph\
    scalar-vacuum configurations of arbitrary dimension in various theories
    of gravity begun in Refs.\,[1--3].          
    We will here consider scalar-tensor theories (STT) belonging to the
    Bergmann-Wagoner-Nordtvedt family, where the Lagrangian depends on two
    essential arbitrary functions of the scalar field. It can be mentioned
    that STT are among the viable alternatives to \GR (GR), and their
    different versions emerge in the field limits of the candidate
    ``theories of everything''.

    The field equations of an arbitrary STT are reduced by a
    conformal mapping to the equations of GR with a scalar field possessing
    a certain potential (the so-called Einstein frame). This paper will pay
    special attention to the properties of such mappings. The point is that,
    when a manifold $\M[g]$ is conformally mapped to another manifold
    $\oM[\og]$ (relating the metrics by $g\mn = F(x)\og\mn$), the global
    properties of both manifolds are the same as long as the conformal
    factor $F$ is everywhere smooth and finite.  It can happen, however,
    that a singular surface in $\oM$ maps to a regular surface \Str\ in $\M$
    due to a singularity in the conformal factor $F$.  Then $\M$ can be
    continued in a regular manner through this surface, and the global
    properties of $\M$ can be considerably richer than those of $\oM$: in
    the new region one can possibly find, e.g., new horizons or another
    spatial infinity. A known example of this phenomenon, to be called {\sl
    conformal continuation,\/} is provided by the properties of the \ssph\
    solution for a conformally coupled scalar field in GR \cite{bbm70, br73}
    as compared with the corresponding solution for a minimally coupled
    scalar field --- see \sect 6.

    The Einstein-frame action for vacuum configurations in STT reads
\beq 						             \label{act-d}
    S_{\rm E} =\int d^D x\,\sqrt{\og}\,[\cR_{\rm E}+ (\d\psi)^2 - 2V(\psi)],
\eeq
    i.e., coincides with the action of GR with a minimally coupled real
    scalar field $\psi$ possessing a potential $V(\psi)$.

    The field equations due to (\ref{act-d}) with nontrivial
    potentials can be integrated explicitly in very few cases, even for
    highly symmetric configurations such as cosmological or \ssph\ ones.
    Nevertheless, rather much can be said about the nature of the solutions.
    Examples of such general statements for nonnegative potentials $V$ are
    the no-hair theorems \cite{bek72} discarding nontrivial scalar field
    for asymptotically flat \bhs\ and the generalized Rosen theorem
    \cite{brsh} claiming that an asymptotically flat solution with a
    positive mass cannot have a regular centre.

    It is also of interest what can happen if the asymptotic flatness and/or
    $V\geq 0$ assumptions are abandoned. Both assumptions are frequently
    violated in modern studies. Negative potential energy densities, in
    particular, the cosmological constant $V=\Lambda < 0$ giving rise to the
    anti--de Sitter (AdS) solution or AdS asymptotic, do not lead to
    catastrophes (if bound below), are often treated in various aspects
    and quite readily appear from quantum effects like vacuum polarization.

    Our previous papers \cite{vac1,vac2} have provided some essential
    restrictions on the possible behaviour of solutions of the theory
    (\ref{act-d}) with arbitrary $V(\psi)$ in $D$ dimensions. It has
    been shown, in particular, that, whatever is the potential and
    irrespective of the asymptotic conditions, the variable scalar field
    adds nothing to the list of causal structures known for $\psi=\const$. In
    the latter case $V$ becomes a cosmological constant, and the
    corresponding exact solutions are well known (\Sch, \Sch-\dS, \Sch-\adS\
    and their multidimensional analogues) along with their causal structures.

    The possibility of regular configurations without a centre (wormholes and
    horns) was also ruled out.

    As was shown in \cite{vac1,vac2}, the above results can be extended to
    (i) generalized scalar field Lagrangians in GR, with an arbitrary
    dependence on the $\psi$ field and its gradient squared, and (ii) to
    multiscalar field theories of sigma-model type in GR. To scalar-tensor
    theories, as is clear from the aforesaid, the same results can be
    extended only partly and only in the absence of conformal continuation
    (CC). This phenomenon is of particular interest since it widens the set
    of possible configurations. A study of possible CCs in the Jordan frames
    of STT was begun in \cite{vac3} and is continued here more
    systematically and in more detail. Moreover, we will discuss the global
    properties of conformally continued space-times.

    We will here avoid a detailed discussion of which conformal frame
    (Jordan or Einstein) in STT should be regarded as the physical one,
    refering to the paper \cite{bm01} and references therein. Only one
    comment is in order: when an STT emerges in a weak-field or low-energy
    limit of some more fundamental theory, its Lagrangian generally contains
    the scalar curvature with a $\phi$-dependent factor, thus leading to one
    of numerous possible Jordan frames (e.g., the string metric in models of
    string origin). So, by origin, it is this formulation of the theory
    that should be used for studying such fundamental issues as topology,
    singularities, causal properties, etc., although a comparison with
    observations may require a different formulation.

    The paper is organized as follows. \sect 2 presents the field equations.
    \sect 3 reviews the known results on scalar vacuum structures in GR
    and configurations described by generic STT solutions. We begin with a
    brief description of purely vacuum structures in $D$-dimensional GR with
    a cosmological constant and then reproduce the no-go theorems of
    Refs.\,\cite{vac1,vac2} on the properties of scalar vacuum in GR and
    mention some other known theorems and examples. Two theorems, providing
    the necessary and sufficient conditions under which a given STT contains
    a CC, are formulated and proved in \sect 4. \sect 5
    discusses the global properties of Jordan-frame space-times in the
    present of CCs. It turns out, in particular, that even the presence of
    CCs does not enlarge the number of possible horizons and hence the above
    list of global causal structures. It is shown that one of generic
    structures created by CCs is a traversable wormhole. The whole
    space-time is then globally regular and static. Some particular kinds of
    singularities can also be created beyond a CC surface. \sect 6
    contains two explicit examples of STT solutions with CCs. One of them
    represents the well-known solution for a conformally coupled scalar
    field in GR, which, in addition to singular cases, contains a family of
    traversable wormhole solutions \cite{br73,viss}. The other is a
    nonsingular model containing an infinite sequence of CCs in
    3-dimensional gravity with a conformally coupled scalar field having a
    certain nonnegative potential.

    To sum up, with all theorems and examples at hand, we now have, even
    without solving the field equations, rather a clear picture of
    what can and what cannot be expected from static scalar-vacuum
    configurations in a general class of STT of gravity with various scalar
    field potentials.

    Throughout the paper all relevant functions are assumed to be
    sufficiently smooth, unless otherwise indicated. The symbol $\sim$, as
    usual, connects quantities of the same order of magnitude. The ends of
    theorem proofs are marked with $\DAL$.

\section {Field equations}      

    The general STT action in a $D$-dimensional pseudo-Riemannian manifold
    $\MJ[g]$ is
\bearr
     S_{\rm STT} = \int d^D x \sqrt{g}                 	     \label{act-J}
		   [f(\phi) \cR
\nnn \inch
		   + h(\phi) (\d\phi)^2 -2U(\phi) + L_m],
\ear
    where $g\mn$ is the metric, $\cR=\cR[g]$ is the scalar curvature,
    $g=|\det g\mn|$, $f$, $h$ and $U$ are functions of the real scalar field
    $\phi$, $(\d\phi)^2 = g\MN\d_\mu\phi\d_\nu\phi$, and $L_m$ is the matter
    Lagrangian. The manifold $\MJ[g]$ with the metric $g\mn$ comprises the
    so-called Jordan conformal frame. The vacuum ($L_m=0$) field equations
    due to (\ref{act-J}) read
\bear   \nq
    \nabla_\alpha(h \nabla^\alpha) \phi
             - \half \, \cR \, f_\phi \eql - dU/d\phi,          \label{SJ}
\yy
	f(\phi) \bigl(\cR\mN - \half \delta\mN \cR\bigr)
    \eql h(\phi) \bigl(-\phi_{,\mu} \phi^{,\nu}
   	+ \half\delta\mN \phi^{,\alpha}\phi_{,\alpha}\bigr)    \label{EE-J}
\nnnv \hspace{-22mm}
      - \delta\mN U(\phi) + (\nabla_{\mu}\nabla^\nu - \delta\mN \DAL)f
      - T\mN{}_{\rm (m)},
\ear
    where $\DAL = \nabla^\alpha \nabla_\alpha$ is the d'Alembert operator,
    $\cR\mN$ is the Ricci tensor, and the last term in (\ref{EE-J}) is the
    energy-momentum tensor of matter. (The usual constant factor $8\pi G$,
    where $G$ is the gravitational constant, can be restored by proper
    re-definition of the variables).

    The standard transition to the Einstein frame, which generalizes
    Wagoner's \cite{wagon} 4-dimensional transformation,
\bear
    g\mn \eql F(\psi) \og\mn,\qquad F=|f|^{-2/(D-2)}, 	      \label{g-wag}
\\
    \frac{d\psi}{d\phi} \eql \pm
    			\frac{\sqrt{|l(\phi)}|}{f(\phi)}, \label{phi-wag}
\qquad
    l(\phi) \eqdef fh + \frac{D{-}1}{D{-}2}\Bigl(\frac{df}{d\phi}\Bigr)^2,
\ear
    removes the nonminimal scalar-tensor coupling express\-ed in the
    $\phi$-dependent coefficient before $\cR$. Putting $L_m=0$ (vacuum), one
    can write the action (\ref{act-J}) in the new manifold $\ME[\og]$ with
    the new metric $\og\mn$ and the new scalar field $\psi$ as follows (up
    to a boundary term):
\bearr \nq
    S_{\rm E} = \int  d^D x \sqrt{\og}             	     \label{act-E}
    \Bigl\{\sign f \bigl[\ocR
        + (\sign l) (\d\psi)^2\bigr] -2V(\psi)\Bigr\},
\nnn
\ear
    where the determinant $\og$, the scalar curvature $\ocR$ and $(\d\psi)^2$
    are calculated using $\og\mn$ and
\beq
       V(\psi) = |f|^{-D/(D-2)} (\psi)\, U(\phi).             \label{UV}
\eeq

    Note that $\sign l = -1$ corresponds to the so-called anomalous STT,
    with a wrong sign of scalar field kinetic energy, while $\sign f=-1$
    means that the effective gravitational constant in the Jordan frame
    (which can be defined as $1/f$ up to a constant factor) is negative. So
    the normal choice of signs is $\sign l = \sign f =1$, when the
    scalar-vacuum action takes the form (\ref{act-d}). We will adhere to
    theories with $l > 0$ in the whole paper, but we shall see that the
    continuations to be discussed in \sect 4--6 lead to $f < 0$
    in some regions of \MJ.

    Among the three functions of $\phi$ entering into (\ref{act-J})
    only two are independent since there is a freedom of transformations
    $\phi = \phi(\phi_{\rm new})$. We assume $h\geq 0$ and use this
    freedom, choosing in what follows%
\footnote
{Another standard parametrization is to put $f(\phi)=\phi$ and $h(\phi)
 = \omega(\phi)/\phi$ (the Brans-Dicke parametrization of the general
 theory (\ref{act-J})).
}
    $h(\phi) \equiv 1$.

    From the viewpoint of the field equations, the transformation
    (\ref{g-wag}), (\ref{phi-wag}) is merely a simplifying substitution.
    Instead of \eqs (\ref{SJ}) and (\ref{EE-J}) (assuming $f>0$), we deal
    in \ME\ with simpler equations due to (\ref{act-d}):
\bear
    \overline{\DAL} \psi + dV/d\psi \eql 0,                     \label{SE}
\\
     \ocR\mN -\half \delta\mN\, \ocR \eql                       \label{EE}
     -\psi_{,\mu}\psi^{,\nu}
                +\half \delta\mN (\d\psi)^2 - \delta\mN V (\psi).
\nnn
\ear
    with the Ricci tensor $\ocR\mN$ and the d'Alembert operator
    $\overline{\DAL}$ corresponding to $\og\mn$.

    Consider \ssph\ configurations, so that the metric
    in \ME\ is written as
\beq                                                           \label{dsE}
     ds_{\rm E}^2 = A(\rho) dt^2 - \frac{d\rho^2}{A(\rho)}
    	        			- r^2(\rho) d\Omega_\od{}^2
\eeq
    where $d\Omega_\od{}^2$ is the linear element on the sphere $\S^{\od}$
    of unit radius, and the scalar field is $\psi=\psi(\rho)$.

    Then \eq (\ref{SE}) and some combinations of the Einstein
    equations (\ref{EE}) have the form
\bear
       (Ar^\od \psi')' \eql r^\od V_\psi;                   \label{phi-d}
\\
       	      (A'r^\od)' \eql - (4/\od)r^\od V;               \label{00d}
\\
     	      \od r''/r  \eql -{\psi'}^2;                     \label{01d}
\\
      A (r^2)'' - r^2 A'' + (\od-2) r' (\lal 2Ar' - A'r)
\nn
     		\eql 2(\od-1); 			    	      \label{02d}
\\
     \od(\od-1)(1-A{r'}^2) - \od A'rr' \eql -Ar^2{\psi'}^2 + 2r^2 V,
                                                            \label{int-d}
\ear
    where the prime denotes $d/d\rho$. Only three of these five equations
    are independent: the scalar equation (\ref{phi-d}) follows from
    the Einstein equations, while \eq (\ref{int-d}) is a first integral of
    the others. Given a potential $V(\psi)$, this is a determined set of
    equations for the unknowns $r,\ A,\ \psi$.

    This choice of the radial coordinate according to the condition
    	$\og_{tt}\og_{\rho\rho} = -1$
    is preferable for considering Killing horizons, which correspond to
    zeros of the function $A(\rho)$, since such zeros are regular points of
    \eqs (\ref{phi-d})--(\ref{int-d}), and therefore one can jointly
    consider regions at both sides of a horizon; moreover, in a close
    neighbourhood of a horizon, the coordinate $\rho$ defined in this way
    varies (up to a positive constant factor) like manifestly well-behaved
    Kruskal-like coordinates used for an analytic continuation of the metric
    \cite{cold}. Therefore this coordinate frame can be called {\it
    quasiglobal\/}.

    The corresponding metric in \MJ\ reads
\bear                                                         \label{dsJ}
    ds_{\rm J}^2
    \eql  F(\psi) \biggl[ A(\rho)dt^2
    		- \frac{d\rho^2}{A(\rho)} - r^2 d\Omega_\od{}^2\biggr]
\nnn\cm
       = \cA(q) dt^2 - \frac{dq^2}{\cA(q)} - R^2 d\Omega_\od{}^2,
\ear
    where we have introduced the quasiglobal coordinate $q$ in \MJ, similar
    to $\rho$ in (\ref{dsE}), such that $g_{tt}g_{\rho\rho} = -1$.
    The quantities in (\ref{dsJ}) and (\ref{dsE}) are related by
\beq
\nq
     \pm dq = F d\rho,\quad \cA(q) = F A(\rho), \quad       \label{q-rho}
       			R(q) = \sqrt{F}r(\rho).
\eeq

    With our convention $h(\phi) \equiv 1$, three independent field
    equations in \MJ\ can be written as follows:
\bear
     f\biggl(\cA_{qq} + \od \cA_q\frac{R_q}{R}\biggr)
     		+ \frac{D}{\od}\cA_q f_q + 2\cA \frac{R_q}{R}f_q \nq
\nn                                                           \label{00q}
	+\frac{2}{\od} \cA f_{qq} + \frac{4}{\od} U \eql 0,
\\
   \od f\frac{R_{qq}}{R}  + \phi_q^2 + f_{qq} \eql 0,           \label{01q}
\\
   \frac{f}{R^2}\biggl[ -(\od-1) + \cA RR_{qq} + \cA R_q^2 \cm
\nn	                                                       \label{02q}
     - \Half R^2 \cA_{qq}
	+\Half(\od-2)R_q(2\cA R_q - R\cA_q)\biggr]
\nn
     	+\biggl(\frac{\cA R_q}{R}-\frac{\cA_q}{2}\biggr)f_q  \eql 0.
\ear
    where the subscript $q$ denotes $d/dq$.

\section {Properties of generic STT solutions}     

\subsection {Some known results for the Einstein frame} 

    Let us enumerate some consequences of \eqs (\ref{phi-d})--(\ref{int-d})
    valid in \ME.

    The first important restriction is the nonexistence of regular
    configurations having no centre ($r=0$), namely, wormholes, horns and
    flux tubes \cite{vac1,vac2}.

    For the metric (\ref{dsE}), a (traversable, Lorentzian) {\it wormhole\/}
    is, by definition, a configuration with two asymptotics at which
    $r(\rho)\to \infty$, hence with $r(\rho)$ having at least one regular
    minimum. A {\it horn\/} is a region where, as $r$ tends to some
    finite value, $\og_{tt} = A$ remains finite whereas the length
    integral $l = \int d\rho/\sqrt{A}$ diverges. In other words, a horn is
    a configuration ending with a regular, infinitely long
    $(\od+1)$-dimensional ``tube'' of finite radius. Such ``horned
    particles'' were discussed as possible remnants of black hole
    evaporation \cite{banks}. Lastly, a {\it flux tube\/} is
    a configuration with $r = \const$, a ``cylindrical'' space.

\Theorem{Theorem 1}
    {The field equations due to (\ref{act-d}) for $D\geq 4$ do not admit
    (i) solutions where the function $r(\rho)$ has a regular minimum,
    (ii) solutions describing a horn, and
    (iii) flux-tube solutions with $\psi\ne\const$.
    }

    The formulation of the theorem and its proof \cite{vac1, vac2}, which
    essentially rests on \eq (\ref{01d}), do not refer to any kind of
    asymptotic, therefore wormhole throats or horns are absent in solutions
    having any large $r$ behaviour --- flat, de Sitter or any other, or
    having no large $r$ asymptotic at all.

    For $D=3$ items (i) and (iii) of Theorem 1 hold, but solutions with a
    horn can exist; though, a horn can only appear at a maximum of
    $r(\rho)$, so that horned configurations have no spatial asymptotic.

    The global causal structure of space-time is unambiguously determined
    (up to identification of isometric surfaces, if any) by the disposition
    of static ($A > 0$) and nonstatic, homogeneous ($A < 0$) regions,
    separated by horizons \cite{walker}--\cite{strobl}. The following two
    theorems severely restrict such possible dispositions.

\Theorem{Theorem 2}
   {Consider solutions of the theory (\ref{act-d}), $D\geq 4$, with the
    metric (\ref{dsE}) and $\psi=\psi(\rho)$. Let there be a static
    region $a < \rho < b \leq \infty$. Then:
\begin{description}\itemsep -2pt
\item [(i)]
    all horizons are simple;
\item [(ii)]
    no horizons exist at $\rho < a$ and at $\rho > b$.
\end{description} }
\vspace{-2ex}
\Theorem {Theorem 2a}
    {A static, circularly symmetric configuration in the theory
    (\ref{act-d}), $D=3$, has either no horizon or one simple horizon.}

    The proof of these theorems \cite{vac1, vac2} employs the properties of
    \eq (\ref{02d}), which can be rewritten in the form
\beq
     r^4 B'' + (\od+2)r^3 r'B' = -2(\od-1)                      \label{02d'}
\eeq
    where $B(\rho) = A/r^2$. This equation shows that $B$ cannot have a
    regular minimum, therefore, having once become negative while moving to
    the left or to the right along the $\rho$ axis, $B(\rho)$ (and hence
    $A(\rho)$) cannot return to zero or positive values.

\Picture{87}
{\unitlength=0.5mm
\special{em:linewidth 0.4pt}
\linethickness{0.4pt}
\begin{picture}(155.00,160.00)(5,-4)
\put(15.00,71.00){\vector(1,0){142.00}}
\put(15.00,3.00){\vector(0,1){160.00}}
\bezier{568}(15.00,111.00)(104.00,114.00)(137.00,155.00)
\bezier{656}(29.00,159.00)(68.00,98.00)(136.00,160.00)
\bezier{568}(25.00,155.00)(46.00,112.00)(140.00,113.00)
\bezier{324}(138.00,148.00)(111.00,121.00)(71.00,106.00)
\bezier{484}(71.00,106.00)(24.00,87.00)(19.00,17.00)
\bezier{840}(145.00,109.00)(33.00,108.00)(21.00,11.00)
\bezier{1268}(24.00,9.00)(69.00,157.00)(137.00,10.00)
\bezier{1096}(30.00,8.00)(65.00,134.00)(133.00,8.00)
\bezier{924}(37.00,6.00)(64.00,111.00)(128.00,7.00)
\put(155.00,111.00){\makebox(0,0)[cc]{1a}}
\put(135.00,115.00){\makebox(0,0)[cb]{1b}}
\put(130.00,157.00){\makebox(0,0)[rb]{2b}}
\put(139.00,155.00){\makebox(0,0)[lc]{2a}}
\put(140.00,145.00){\makebox(0,0)[cc]{3}}
\put(140.00,106.00){\makebox(0,0)[ct]{4}}
\put(145.00,24.00){\makebox(0,0)[lb]{5b}}
\put(143.00,8.00){\makebox(0,0)[lb]{5a}}
\put(137.00,5.00){\makebox(0,0)[lc]{6a}}
\put(132.00,1.00){\makebox(0,0)[cb]{6b}}
\put(124.00,8.00){\makebox(0,0)[rb]{6c}}
\bezier{316}(81.00,105.00)(30.00,126.00)(20.00,148.00)
\bezier{460}(81.00,105.00)(123.00,86.00)(144.00,20.00)
\bezier{788}(15.00,111.00)(109.00,111.00)(141.00,13.00)
\put(151.00,111.00){\line(-1,0){135.91}}
\put(11.00,111.00){\makebox(0,0)[cc]{1}}
\end{picture}
}
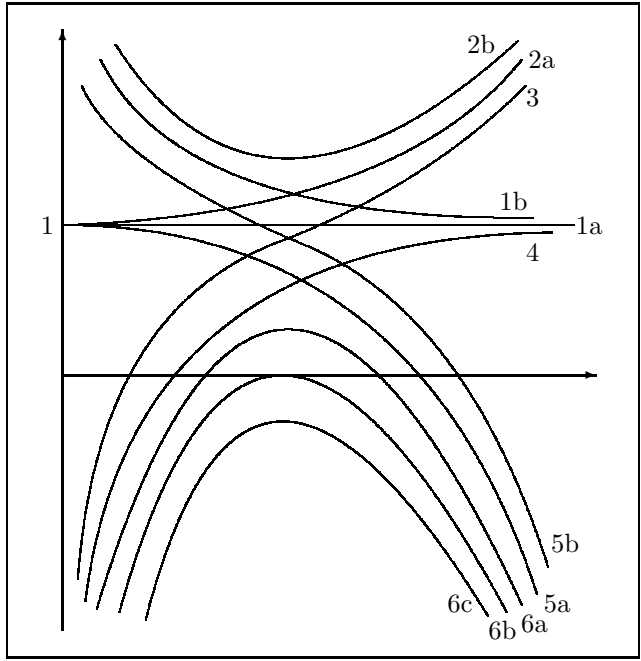
{The behaviour of $A(r)$, \eq(\ref{A-SdS}), for different values of $m$ and
	$\Lambda$.}

\Picture{100}
{\unitlength=0.75mm
\special{em:linewidth 0.4pt}
\linethickness{0.4pt}
\begin{picture}(130.00,136.00)(23,8)
\put(35.00,134.00){\line(1,-1){10.00}}
\put(45.00,124.00){\line(-1,-1){10.00}}
\put(35.00,114.00){\line(0,1){21.00}}
\put(35.00,135.00){\line(1,-1){11.00}}
\put(46.00,124.00){\line(-1,-1){11.00}}
\put(35.00,113.00){\line(0,1){1.00}}
\bezier{120}(65.00,135.00)(55.00,124.00)(65.00,113.00)
\bezier{120}(65.00,135.00)(75.00,124.00)(65.00,113.00)
\put(95.00,135.00){\line(0,-1){21.00}}
\put(95.00,114.00){\line(1,0){20.00}}
\put(115.00,114.00){\line(0,1){21.00}}
\put(115.00,135.00){\line(-1,0){20.00}}
\put(95.00,135.00){\line(1,-1){20.00}}
\put(115.00,115.00){\line(1,-1){1.00}}
\put(116.00,114.00){\line(0,1){21.00}}
\put(116.00,135.00){\line(-1,0){1.00}}
\put(115.00,135.00){\line(-1,-1){20.00}}
\put(95.00,115.00){\line(-1,-1){1.00}}
\put(94.00,114.00){\line(0,1){21.00}}
\put(94.00,135.00){\line(1,0){1.00}}
\put(30.00,82.00){\line(1,1){10.00}}
\put(40.00,92.00){\line(1,0){20.00}}
\put(60.00,92.00){\line(1,-1){10.00}}
\put(70.00,82.00){\line(-1,-1){10.00}}
\put(60.00,72.00){\line(-1,0){20.00}}
\put(40.00,72.00){\line(-1,1){10.00}}
\put(39.00,72.00){\line(-1,1){10.00}}
\put(29.00,82.00){\line(1,1){10.00}}
\put(61.00,92.00){\line(1,-1){10.00}}
\put(71.00,82.00){\line(-1,-1){10.00}}
\put(40.00,72.00){\line(1,1){20.00}}
\put(40.00,92.00){\line(1,-1){20.00}}
\put(95.00,92.00){\line(1,0){20.00}}
\put(115.00,92.00){\line(0,-1){20.00}}
\put(115.00,72.00){\line(-1,0){20.00}}
\put(95.00,72.00){\line(0,1){20.00}}
\put(95.00,92.00){\line(1,-1){20.00}}
\put(115.00,72.00){\line(0,-1){1.00}}
\put(115.00,71.00){\line(-1,0){20.00}}
\put(95.00,71.00){\line(0,1){1.00}}
\put(95.00,72.00){\line(1,1){20.00}}
\put(115.00,92.00){\line(0,1){1.00}}
\put(115.00,93.00){\line(-1,0){20.00}}
\put(95.00,93.00){\line(0,-1){1.00}}
\bezier{136}(65.00,136.00)(77.00,124.00)(65.00,112.00)
\put(23.00,48.00){\line(1,0){49.00}}
\put(22.00,28.00){\line(1,0){52.00}}
\put(71.00,33.00){\line(-1,-1){5.00}}
\put(66.00,28.00){\line(-1,1){20.00}}
\put(46.00,48.00){\line(-1,-1){20.00}}
\put(26.00,28.00){\line(-1,1){4.00}}
\put(22.00,44.00){\line(1,1){4.00}}
\put(26.00,48.00){\line(1,-1){20.00}}
\put(46.00,28.00){\line(1,1){20.00}}
\put(66.00,48.00){\line(1,-1){4.00}}
\put(46.00,49.00){\line(1,0){20.00}}
\put(26.00,49.00){\line(-1,0){4.00}}
\put(22.00,27.00){\line(1,0){4.00}}
\put(46.00,27.00){\line(1,0){20.00}}
\put(87.00,50.00){\line(1,0){43.00}}
\put(130.00,40.00){\line(-1,0){42.00}}
\put(88.00,44.00){\line(1,-1){4.00}}
\put(92.00,40.00){\line(1,1){10.00}}
\put(102.00,50.00){\line(1,-1){10.00}}
\put(112.00,40.00){\line(1,1){10.00}}
\put(122.00,50.00){\line(1,-1){8.00}}
\put(130.00,51.00){\line(-1,0){43.00}}
\bezier{104}(98.00,21.00)(109.00,29.00)(119.00,21.00)
\bezier{120}(97.00,21.00)(109.00,31.00)(120.00,21.00)
\bezier{96}(98.00,21.00)(109.00,15.00)(119.00,21.00)
\put(38.00,107.00){\makebox(0,0)[cc]{1a, 1b}}
\put(65.00,107.00){\makebox(0,0)[cc]{2a, 2b}}
\put(105.00,107.00){\makebox(0,0)[cc]{3}}
\put(50.00,64.00){\makebox(0,0)[cc]{4}}
\put(105.00,64.00){\makebox(0,0)[cc]{5a, 5b}}
\put(45.00,18.00){\makebox(0,0)[cc]{6a}}
\put(98.00,34.00){\makebox(0,0)[cc]{6b}}
\put(98.00,14.00){\makebox(0,0)[cc]{6c}}
\put(39.00,124.00){\makebox(0,0)[cc]{R}}
\put(65.00,124.00){\makebox(0,0)[cc]{R}}
\put(99.00,125.00){\makebox(0,0)[cc]{R}}
\put(111.00,125.00){\makebox(0,0)[cc]{R}}
\put(40.00,82.00){\makebox(0,0)[cc]{R}}
\put(60.00,82.00){\makebox(0,0)[cc]{R}}
\put(99.00,82.00){\makebox(0,0)[cc]{R}}
\put(112.00,82.00){\makebox(0,0)[cc]{R}}
\put(26.00,38.00){\makebox(0,0)[cc]{R}}
\put(46.00,38.00){\makebox(0,0)[cc]{R}}
\put(66.00,38.00){\makebox(0,0)[cc]{R}}
\put(50.00,88.00){\makebox(0,0)[cc]{T${}_+$}}
\put(105.00,118.00){\makebox(0,0)[cc]{T${}_+$}}
\put(105.00,88.00){\makebox(0,0)[cc]{T${}_+$}}
\put(36.00,32.00){\makebox(0,0)[cc]{T${}_+$}}
\put(56.00,44.00){\makebox(0,0)[cc]{T${}_+$}}
\put(92.00,46.00){\makebox(0,0)[cc]{T${}_+$}}
\put(112.00,46.00){\makebox(0,0)[cc]{T${}_+$}}
\put(102.00,43.00){\makebox(0,0)[cc]{T${}_+$}}
\put(122.00,43.00){\makebox(0,0)[cc]{T${}_+$}}
\put(50.00,75.00){\makebox(0,0)[cc]{T${}_-$}}
\put(105.00,75.00){\makebox(0,0)[cc]{T${}_-$}}
\put(105.00,131.00){\makebox(0,0)[cc]{T${}_-$}}
\put(36.00,44.00){\makebox(0,0)[cc]{T${}_-$}}
\put(56.00,31.00){\makebox(0,0)[cc]{T${}_-$}}
\put(109.00,21.00){\makebox(0,0)[cc]{T${}_+$}}
\end{picture}
}
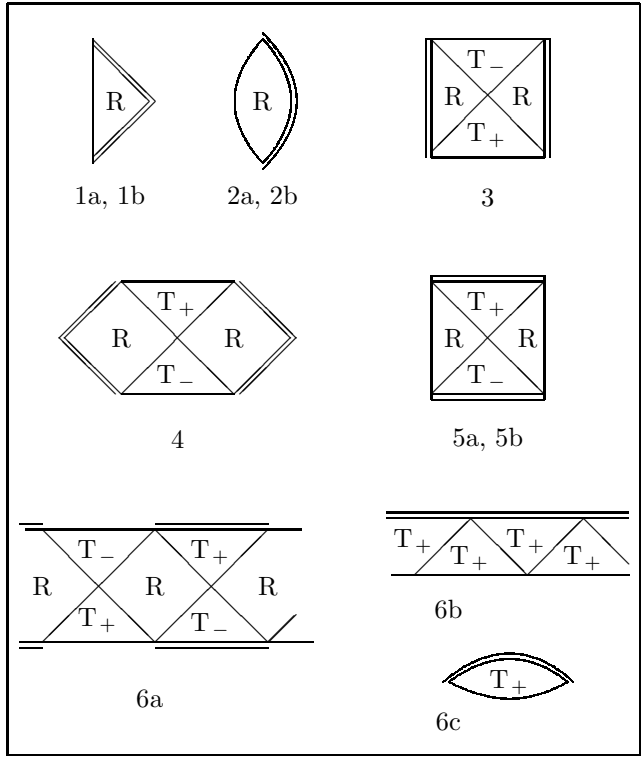
{Carter-Penrose diagrams for different cases of the metric (\ref{g-SdS}),
	(\ref{A-SdS}), labelled according to \fig 1. The R and T
	letters correspond to R and T space-time regions; T${}_+$ and
	T${}_-$ denote expanding and contracting T region (i.e., with $r$
	increasing and decreasing with time, respectively). Single
	lines on the border of the diagrams denote $r=0$, double lines ---
	$r=\infty$. Diagrams 6b and 6c are drawn for the case of expanding
	Kantowski-Sachs cosmologies; to obtain diagrams for contracting
	models, one should merely interchange $r=0$ and $r=\infty$ and
	replace T${}_+$ with T${}_-$. Diagrams 6a and 6b admit
	identification of isometric timelike sections.}

    Theorems 2 and 2a show that the possible disposition of zeros of the
    function $A(\rho)$ is the same as in the case of vacuum with a
    cosmological constant. Therefore the list of possible global
    causal structures is also the same.

    Let us, for reference purposes, enumerate these structures. The metric
    satisfying \eqs (\ref{00d})--(\ref{int-d}) with $\psi'=0$,
    $V = \Lambda = \const$ is
\beq                                                          \label{g-SdS}
    ds^2 = A(r) dt^2 - \frac{dr^2}{A(r)} - r^2\,d\Omega_\od{}^2
\eeq
    [it is (\ref{dsE}) with $\rho \equiv r$] where
\beq                                                          \label{A-SdS}
    A(r) = 1 - \frac{2m}{r^{\od-1}} - \frac{2\Lambda r^2}{\od(\od+1)}.
\eeq
    This is the multidimensional \Sch-\dS\ solution. Its special cases
    correspond to the \Sch\ ($\od=2$, $\Lambda=0$) and Tangherlini (any
    $\od$, $\Lambda=0$) solutions and the \dS\ solution in arbitrary
    dimension when $m=0$, called \adS\ (AdS) in case $\Lambda <0$.

    Different qualitative behaviours of $A(r)$ for different values of
    $\Lambda$ and $m$ correspond to the following structures \cite{SdS}:
\begin{enumerate}\itemsep -1.5pt
\item
    $\Lambda = 0,\ m\leq 0$: curves 1a and 1b in \fig 1, diagram 1 in \fig 2
    (Minkowski and $m<0$ \Sch, respectively).
\item
    $\Lambda < 0,\ m\leq 0$: curves 2a and 2b in \fig 1, diagram 2 in \fig 2
    (AdS and $m<0$ \Sch-AdS).
\item
    $\Lambda < 0,\ m > 0$: curve 3 in \fig 1, diagram 3 in \fig 2
    (\Sch-AdS).
\item
    $\Lambda = 0,\ m > 0$: curve 4 in \fig 1, diagram 4 in \fig 2 (\Sch).
\item
    $\Lambda > 0,\ m\leq 0$: curves 5a and 5b in \fig 1, diagram 5 in \fig 2
    (\dS\ and $m<0$ \Sch-\dS).
\item
    $\Lambda > 0,\ m > 0$: curves 6a, 6b and 6c in \fig 1, and the
    corresponding diagrams in \fig 2 \ (\Sch-\dS\ in case 6a and
    Kantowski-Sachs homogeneous cosmologies in cases 6b and 6c).
\end{enumerate}
    The centre $r=0$ is regular for $m=0$ and singular for $m\ne 0$.

    In case 6, given a particular value of $\Lambda > 0$, the solution
    behaviour depends on the mass parameter $m$. When $m$ is smaller than
    the critical value
\beq                                                          \label{mcrit}
 	m_{\rm cr} = \frac{1}{\od+1}
		\biggl[ \frac{\od(\od-1)}{2\Lambda}\biggr]^{(\od-1)/2},
\eeq
    there are two horizons, the smaller one being interpreted as a \bh\
    horizon and the greater one as a cosmological horizon. If $m=m_{\rm
    cr}$, the two horizons merge, and one has two homogeneous T regions
    separated by a double horizon. Lastly, the solution with
    $m > m_{\rm cr}$ is purely cosmological and has no Killing horizon.

    In (2+1)-dimensional gravity ($\od=1$), according to Theorem 2a, the
    list is even shorter: the structures corresponding to the curves 6a and
    6b are absent.

    Let us also mention, for completeness, some results known for $D=4$ and
    most probably admitting a generalization to other dimensions.

\def\parag#1{\medskip\noi{\bf #1}}
\parag{No-hair theorems} state that (1) if $V\geq 0$, an asymptotically flat
    \bh\ cannot have a nontrivial scalar field \cite{bek72, bek98, ad-pier};
    (2) if $V\geq 0$ and $d^2 V/d\psi^2 \geq 0$ (a convex potential), an
    asymptotically \adS\ \bh\ cannot have a nontrivial scalar field
    \cite{to-ma99}.

\parag{The generalized Rosen theorem} states that, provided $V\geq 0$,
    a particlelike solution with a regular centre, a flat asymptotic and
    positive mass does not exist \cite{brsh}.

\medskip
    These theorems cannot be directly extended to STT in the Jordan frame
    and will not be discussed any more, though an attempt to formulate
    additional conditions able to provide such extensions may be of
    interest.

\parag{Explicit examples} have been obtained, confirming the existence of
    some kinds of solutions admitted by the above theorems. Thus, there
    exist: (1) \bhs\ possessing nontrivial scalar fields (scalar hair), with
    $V\geq 0$, but with non-flat and non-de Sitter asymptotics
    \cite{Mann95}; (2) \bhs\ with scalar hair and flat asymptotics, but
    partly negative potentials \cite{vac2}; (3) configurations with a
    regular centre, a flat asymptotic and positive mass, but also with
    partly negative potentials \cite{vac2}.

    Thus black holes with scalar hair are not excluded in general, but such
    objects as regular black holes \cite{dym}, possessing a regular centre
    and a global structure coinciding with that of Reissner-Nordstrom or
    Reissner-Nordstrom-de Sitter space-time, are ruled out.

\subsection {Generic solutions in the Jordan frame}    

    It should be, above all, noted that when a space-time manifold $\ME[\og]$
    (the Einstein frame) with the metric (\ref{dsE}) is conformally
    mapped into another manifold $\MJ[g]$ (the Jordan frame), equipped with
    the same coordinates, according to the law
\beq
    g\mn = F(\rho) \og\mn,                                \label{g-conf}
\eeq
    it is easily verified that a horizon $\rho=h$ in \MJ\ passes into a
    horizon of the same order in \ME, a centre ($r=0$) and an
    asymptotic ($r\to \infty$) in \MJ\ pass into a centre and an asymptotic,
    respectively, in \ME\ if the conformal factor $F(\rho)$ is regular
    (i.e., finite, at least C${}^2$-smooth and positive) at the
    corresponding values of $\rho$. A regular centre passes to a regular
    centre and a flat asymptotic to a flat asymptotic under evident
    additional requirements.

    The validity of Theorems 1, 2 and 2a in the Jordan frame
    depends on the nature of the conformal mapping (\ref{g-wag})
    that connects $\MJ[g]$ with $\ME [\og]$). There are four variants:
\begin{description}\itemsep -2pt
\item[I.]
	$\MJ \ \map\ \ME$,
\item[II.]
	$\MJ \ \map\ (\ME' \subset \ME$),
\item[III.]
	$(\MJ' \subset \MJ) \ \map\ \ME$,
\item[IV.]
	$(\MJ' \subset \MJ) \ \map\ (\ME' \subset \ME$),
\end{description}
    where $\map$ is a diffeomorphism preserving the metric signature.
    The last three variants are possible if the conformal factor $F$
    vanishes or blows up at some values of $\rho$, which then mark the
    boundary of $\MJ'$ or $\ME'$.

    A situation of the kind III or IV can be called a {\it conformal
    continuation\/} (CC) from \ME\ into \MJ.

    One can notice that such continuations can only occur for special
    solutions: to admit a CC, the singularity in \ME\ should be removable
    by a conformal factor, i.e., be, in a sense, isotropic. Moreover, the
    factor $F$ should have precisely the behaviour needed to remove it.

    Thus generic situations are I and II, the latter meaning that the
    factor $F$ ``spoils'' the geometry and creates a singularity.
    In these cases Theorem 2 (or 2a for $D=3$) on horizon dispositions is
    obviously valid in \MJ.

    The manifolds \MJ\ then cannot have other causal structures than those
    depicted in \fig 2. This is manifestly true for STT with $f(\phi) >0$.

    Theorem 1 cannot be directly transferred to \MJ\ in any case except the
    trivial one, $F=\const$. In particular, minima of $g_{\theta\theta}$
    (wormhole throats) can appear. It is only possible to assert without
    specifying $F(\psi)$ that wormholes as global entities are impossible
    in \MJ\ in case I if the conformal factor $F$ is finite in the whole
    range of $\rho$, including the boundary values. Indeed, if we suppose
    that there is such a wormhole, it will immediately follow that there are
    two large $r$ asymptotics and a minimum of $r(\rho)$ between them even
    in \ME, in contrast to Theorem 1 which is valid there. Wormholes are
    also absent in case II since we then have a singularity instead of at
    least one of the asymptotics.

    The above-mentioned examples of \bhs\ with scalar hair when the
    potential is not positive-definite or the asymptotic is non-flat are
    also directly transferred to \MJ\ provided $F(\psi)$ is regular at
    least outside the horizon. Given a particlelike solution in \ME, with
    a regular centre and positive mass, the condition that a solution with
    similar properties occurs in \MJ\ can also be easily formulated, but we
    will not concentrate on this question here.

    Conformal continuations, if any, can in principle lead to other,
    maybe more complex structures. In what follows we will try to answer two
    questions: (1) under which conditions the mapping (\ref{g-conf}) creates
    a conformal continuation in STT and (2) what can be the nature
    of conformally continued solutions in the Jordan frame.

\section{Conformal continuation conditions}              

\subsection{Preliminaries}

    A CC from \ME\ into \MJ\ can occur at such values of the scalar field
    $\phi$ that the conformal factor $F$ in the mapping (\ref{g-wag}) is
    singular while the functions $f$, $h$ and $U$ in the action
    (\ref{act-J}) are regular.
    This means that at $\phi=\phi_0$, corresponding to a
    possible transition surface \Str, the function $f(\phi)$ has a zero of a
    certain order $n$. We then have in the transformation (\ref{phi-wag})
    near $\phi=\phi_0$ in the leading order of magnitude
\beq
    f(\phi) \sim \Df^n, \qquad   n = 1,2,\ldots,
    	 \qquad		   \Df \equiv  \phi-\phi_0.      \label{phi0}
\eeq

    One can notice, however, that $n > 1$ leads to $l(\phi_0) = 0$
    (recall that by our convention $h(\phi)\equiv 1$). This generically
    leads to a curvature singularity in \MJ, as can be seen from the
    trace of \eqs (\ref{EE-J}):
\bearr
      l(\phi) \cR = - \biggl(1 + 2 \frac{D-1}{D-2}\,f_{\phi\phi}\biggr)
      		\phi^\alpha \phi_\alpha
\nnn \cm \cm                                                   \label{trace}
	 + \frac{2}{D-2}\biggl[DU + (D-1)f_\phi U_\phi \biggr],
\ear
    (the subscript $\phi$ denotes $d/d\phi$).
    If the right-hand side of (\ref{trace}) is nonzero at $\phi=\phi_0$ at
    which $l=0$, the scalar curvature $\cR$ is infinite. There can be
    special choices of $f$ and $U$ such that this singularity is avoided,
    but we will ignore this possibility and simply assume $l > 0$ at \Str.

    Thus, according to (\ref{phi-wag}), we have near \Str ($\phi=\phi_0$):
\bear
    f (\phi) \sim \Df \sim \e^{-\psi\sqrt{\od/(\od+1)}},       \label{Df}
\ear
    where without loss of generality we choose the sign of $\psi$ so that
    $\psi\to\infty$ as $\Df \to 0$.

    In the CC case, the metric $\og\mn$ specified by (\ref{dsE}) is singular
    on \Str\ while $g\mn = F(\psi)\og\mn$ is regular. There are two
    opportunities. The first one, to be called CC-I for short, is that
    \Str\ is an {\it ordinary regular surface\/} in \MJ, where both $g_{tt}=
    \cA = FA$ and $-g_{\theta\theta} = R^2 = Fr^2$ (squared radius of \Str)
    are finite.  (Here $\theta$ is one of the angles that parametrize the
    sphere $\S^{\od}$.) The second variant, to be called CC-II, is that
    \Str\ is a {\it horizon} in \MJ. In the latter case only
    $g_{\theta\theta}$ is finite, while $g_{tt}=0$. We will consider these
    two kinds of CC separately.

    In both cases some necessary conditions for CC are easily obtained
    using the field equations in \ME, but these equations describe the
    system only on one side of \Str. Another Einstein frame may be built for
    the region beyond \Str, but in this case there arises the problem of
    matching the solutions obtained in two non-intersecting regions.
    Therefore to prove sufficient conditions for the existence of solutions
    in \MJ\ which are regular on and near \Str\ we have to deal with the
    field equations in \MJ.

\subsection {Continuation through an ordinary sphere (CC-I)}

    Given a metric $\og\mn$ of the form (\ref{dsE}) in \ME, a CC-I
    can occur if
\beq                                                          \label{CC-g}
    F(\psi) = |f|^{-2/\od} \sim 1/r^2 \sim 1/A
\eeq
    as $\psi\to\infty$, while the behaviour of $f$ is specified by
    (\ref{Df}). The surface \Str, being regular in the Jordan frame, is
    singular in \ME\ ($r^2 \sim A \to 0$): it is either a singular centre
    if the continuation occurs in an R-region, or a cosmological singularity
    in the case of a T-region.

    The following theorem is valid:

\Theorem{Theorem 3}
    {Consider scalar-vacuum configurations with the metric (\ref{dsJ})
    and $\phi=\phi(q)$ in the theory (\ref{act-J}) with $h(\phi) \equiv 1$
    and $l(\phi) >0$. Suppose that $f(\phi)$ has a simple zero at some
    $\phi = \phi_0$, and $|U(\phi_0)| < \infty$. Then:
\Description{
\item[(i)]
    there exists a solution in \MJ, smooth in a neighbourhood of the surface
    \Str\ ($\phi=\phi_0$), which is an ordinary regular surface in \MJ;
\item[(ii)]
    in this solution the ranges of $\phi$ are different on different sides
    of \Str.  }}

\noi{\bf Proof.}
    Let us begin with item (ii): given a CC-I, we will show that
    $d\phi/dq\ne 0$ at $\phi=\phi_0$, so that $\phi_0$ is not a maximum or
    minimum of $\phi (q)$.

    Indeed, it can be deduced from the conditions (\ref{Df}) and
    (\ref{CC-g}) and \eq (\ref{01d}) that near \Str
\beq
      r(\rho) \sim (\rho-\rho_0)^{1/D}                       \label{r_trans}
\eeq
    where $\rho=\rho_0$ is the location of \Str. It then follows that
    $q_0=q(\rho_0)$ is finite on \Str\ and both $\Df$ and $q-q_0$ behave as
    $r^\od$ in its neighbourhood, hence $d\phi/dq$ is finite.

    With this necessary condition, we can prove item (i), seeking a
    solution to \eqs (\ref{00q})--(\ref{02q}) in an appropriate form.
    Here, the unknowns are $\cA(q),\ R(q),\ \phi(q)$, while
    $f(\phi)$ and $U(\phi)$ are prescribed by the choice of the theory.
    However, since $f(\phi_0) =0$ and $df/d\phi \ne 0$ at $\phi=\phi_0$, we
    can treat $\phi(f)$ as a known function in a certain neighbourhood of
    $f=0$ and consider $f(q)$ as an unknown instead of $\phi(q)$.

    Let \Str\ be located (without loss of generality) at $q=0$.
    It is sufficient to find a solution in the form of a power series in $q$
    near $q=0$. Since $R$ and $A$ should be finite at \Str, while $f=0$ and
    $df/dq \ne 0$, we seek a solution in the form
\bear  \nhq
      \cA(q) \eql \sum\nolimits_{n=0}^{\infty} A_n q^n/n!
                      = A_0 + A_1 q + \half A_2 q^2 + \ldots,
\nn \nhq
	R(q) \eql \sum\nolimits_{n=0}^{\infty} R_n q^n/n!
		      = R_0 + R_1 q + \half R_2 q^2 + \ldots,
\nn \nhq                                                        \label{qn}
	f(q) \eql \sum\nolimits_{n=1}^{\infty} f_n q^n/n!
		      = f_1 q + \half f_2 q^2 + \ldots,
\ear
    with nonzero $A_0,\ R_0,\ f_1$.

    Substituting (\ref{qn}) into the equations, we see that in the senior
    order of magnitude, $O(1)$, the coefficients $A_1,\ R_1,\ f_2$ are
    expressed in terms of $A_0,\ R_0,\ f_1$ and $U(\phi_0)$. Next powers of
    $q$ express the further expansion factors in terms of the previous
    ones. Namely, in every order of magnitude $O(q^n)$, $n>0$, \eq
    (\ref{01q}) gives
\beq
	(n-1) \od R_n/R_0 + f_{n+1} /f_1 = \ldots,           \label{f_to_R}
\eeq
    where the dots on the right mean various combinations of coefficients
    of the previous orders as well as power expansion factors of the known
    functions $U(\phi)$ and $\phi(f)$. Then, excluding $f_{n+1}$ from the
    other two equations in the order $O(q^n)$, we obtain a set of two linear
    algebraic equations for $A_n/A_0$ and $R_n/R_0$:
\bear
	A_n/A_0 - 2 R_n/R_0 \eql \ldots,                        \label{n3a}
\nn
	(n\od +2) A_n/A_0 -2\od(n-2) R_n/R_0 \eql \ldots,
\ear
    whose determinant is equal to $4(\od+1)$ for any $n$. We conclude that
    all the expansion factors in (\ref{qn}) are uniquely expressed in
    terms of $A_0,\ R_0,\ f_1$ and the expansion factors of the known
    functions. This proves the existence of the solution in \MJ\ near \Str.
    $\DAL$

\medskip
    The order of smoothness of the solution obtained depends on the
    smoothness of the original functions $f,\ h,\ U$. If they are
    $C^{\infty}$, as is natural for a field theory, then the metric
    functions and $\phi$ are also $C^\infty$.

    The existence of such a solution automatically implies the existence of
    the corresponding solutions on different sides of \Str\ in two different
    Einstein frames. These solutions are special, being restricted by \eq
    (\ref{CC-g}). As follows from the proof, the solution in \MJ, and hence
    its counterparts in both \ME, contain two essential integration
    constants ($R_0$ and $f_1$, whereas $A_0$ determines the time scale on
    \Str\ and can be chosen arbitrarily).

    It is of interest that, under the CC-I conditions, the potential
    $V(\psi)$ in \ME\ (although it may even blow up) is inessential:  the
    solution is close to Fisher's scalar-vacuum solution \cite{fisher} for
    $D=4$ or its modification in other dimensions.

    In case $D=3$, as follows from \eq (\ref{02d}), a necessary condition
    for CC is $A/r^2 = \const$.

    One can also notice that no restriction other than regularity is imposed
    on the potential $U$,%
\footnote
{The conclusion that $U(\phi)=0$ on \Str, obtained in \cite{vac3},
appeared there due to an additional assumption on the form of
the expansion of $A(\rho)$ in powers of $\rho$. An inspection shows
that this assumption is unnecessary.
}
    in particular, $U$ may vanish in some region or in the whole space. The
    latter case will be used as an explicit example of CC in \sect 6.1.

\subsection {Continuation through a horizon in \MJ\ (CC-II)}

    Let us suppose that in the metric (\ref{dsJ}) a certain value of $q$
    (without loss of generality, $q=0$) corresponds to a horizon of order
    $k\geq 1$. This means that $q=0$ is a zero of order $k$ of the function
    $\cA(q)$.

    Suppose now that this horizon is \Str, a transition sphere in a CC. In
    other words, in the vicinity of $q=0$, $f(\phi) \sim \Df$. One can
    directly verify that in the corresponding value $\rho_0$ of the
    coordinate $\rho$ in the Einstein-frame metric (\ref{dsE}) is finite,
    and we can choose for convenience $\rho_0 = 0$. We thus have near $q=0$
\beq
    \cA(q) = AF \sim q^k, \qquad  R^2(q) = Fr^2 = O(1).    \label{horiz}
\eeq

    Now, the question is: under which requirements to the original theory
    a horizon in \MJ, described by (\ref{horiz}), can be a transition sphere
    \Str. An answer is given by the following theorem.

\Theorem{Theorem 4}
    {Consider scalar-vacuum configurations with the metric (\ref{dsJ})
    and $\phi=\phi(q)$ in the theory (\ref{act-J}) with $h(\phi) \equiv 1$
    and $l(\phi) >0$. Suppose that $f(\phi)$ has a simple zero at some
    $\phi = \phi_0$.
    There exists a solution in \MJ, smooth in a neighbourhood of the
    surface \Str\ ($\phi=\phi_0$), which is a Killing horizon in \MJ,
    {\bf if and only if:}
\Description{
\item[(a)]  $D\geq 4$,
\item[(b)]  $\phi_0$ is a simple zero of $U(\phi)$,
\item[(c)]  $dU/df >0$ at $\phi=\phi_0$.
    }
    Then, in addition,
\Description{
\item[(d)]
     \Str\ is a second-order horizon, connecting two T-regions in \MJ;
\item[(e)]
     the ranges of $\phi$ are different on the two sides of \Str.
    }}

\noi{\bf Proof.}
    {\bf Necessity.}
    Given a CC-II, we will prove items (a)--(e).

    Let us use \eqs (\ref{00d})--(\ref{02d}) in \ME. In particular, \eq
    (\ref{02d}), which contains only $A(\rho)$ and $r(\rho)$,
    can be rewritten in the form
\beq
    \bigl [ r^D(A/r^2)'\bigr]' + 2 (\od-1)r^{\od-2} =0.        \label{02'}
\eeq

    Suppose CC-II at $\rho=0$ ($q=0$) which is a horizon of order $k$
    in \MJ. Let us put $\od > 1$ and assume for certainty, without
    generality loss, that $q > 0$ as $\rho\to +0$. We have
    $dq/d\rho \sim F \sim 1/r^2$ and $A/r^2 \sim q^k$. Therefore the first
    term in (\ref{02'}) at small $\rho$ behaves as
\[
	(\sign A)\, [r^\od q^{k-1}]'.
\]
    Since both $r(\rho)$ and $q$ are growing functions of $\rho$,
    this derivative is nonnegative, whereas the second term in (\ref{02'})
    is manifestly positive. The only way to satisfy (\ref{02'}) is to
    put $A<0$. In other words, the horizon is approached from a T-region,
    where we deal with a Kantowski-Sachs type cosmological model.

    Such a reasoning applies to approaching the surface $q=0$ from either
    side, therefore the horizon \Str\ connects two T-regions
    and is thus of even order.

    We must also ascertain that the orders of magnitude of the two terms in
    (\ref{02'}) are the same. This is only true if $k=2$, as can be easily
    verified using (\ref{r_trans}), which now reads $r\sim \rho^{1/D}$.
    So item (d) is proved.

    \eqs (\ref{Df}) and (\ref{r_trans}) can be used to show that the
    derivative $d\phi/dq$ is finite at $q=0$, leading to item (e).

    The behaviour of $U(\phi)$ can be determined using \eq (\ref{00d}),
    taking into account the relation (\ref{UV}) between $U$ and $V$
    and the conditions (\ref{horiz}) with $k=2$. We find in this way that
    $U(\phi) \sim \Df$ (i.e., the potential has a first-order zero)
    and that $dU/df > 0$ at $\phi=\phi_0$, so items (b) and (c) hold.

    It remains to rule out $\od=1$ (3-dimensional gravity). In this case
    (\ref{02'}) leads to $(A/r^2) = c_1/r^3$, $c_1=\const$. The value
    $c_1=0$ is excluded since we must have $A/r^2 \sim q^k$. For $c_1\ne 0$
    we obtain $(q^k)' = kq^{k-1} dq/d\rho$, whence due to (\ref{horiz})
    $q^{k-1} \sim 1/r \to \infty$, whereas $q \to 0$ and $k\geq 1$.
    This contradiction proves item (a).

\medskip\noi
   {\bf Sufficiency.}
    As in Theorem 3, the existence of a solution in \ME, smoothly continued
    across \Str, is proved using \eqs (\ref{00q})--(\ref{02q}) and a
    power expansion for a sought-for solution. It is again
    convenient to treat $\phi(f)$ as a known function and $f(q)$ as an
    unknown, so again the unknowns are $\cA(q),\ R(q)$ and $f(q)$.

    Seeking solutions to \eqs (\ref{00q})--(\ref{02q}) as series in $q$,
    we use again the expansions (\ref{qn}), but, under the present necessary
    conditions, we put there $A_0 = A_1 =0$ and suppose
    nonzero $A_2,\ R_0,\ f_1$. With these expansions substituted, the
    equations again lead to a chain of recurrent relations for the
    coefficients, slightly more involved than in Theorem 3.

    In the orders $O(1)$ and $O(q)$, \eq (\ref{00q}) leads to
\[
    U(\phi_0)=0, \cm (\od+D) A_2 + 4 U_{f0} =0
\]
    (where $U_{f0} = dU/df\big|_{f=0}$), hence $U_{f0} >0$, in
    agreement with the necessary conditions. \eq (\ref{01q}) [$O(1)$]
    expresses $f_2$ in terms of $f_1$ and $d\phi/df (f=0)$. \eq(\ref{02q})
    [$O(1)$] gives $A_2 = - (\od-1)/R_0^2$.  Furthermore, \eqs
    (\ref{00q}) [$O(q^2)$] and (\ref{02q}) [$O(q)$] yield two equations
    determining $R_1$ and $A_3$:
\bear
     (3\od+2) A_3/A_2 + 2\od(\od+1) R_1/R_0 \eql \ldots,      \label{A3R1}
\nn
     3A_3/A_2 + 2(\od-1) R_1/R_0 \eql \ldots,
\ear
    where the dots on the right replace combinations of previously known
    constants. The determinant of this set of linear equations with respect
    to $A_3/A_2$ and $R_1/R_0$ is $-4(2\od + 1) <0$ for $\od >0$. We
    thus know the coefficients up to $R_1,\ f_2,\ A_3$. Further coefficients
    are determined recursively from further orders of magnitude in all three
    equations:
\bear
      (n+D/\od)a_n + n(n-1)(\od+1)b_n \cm
\nn
		       	+ (n-1)(n+D)c_n/D \eql \ldots,
\nn
      (n+2) \od b_n + c_n \eql \ldots,                          \label{abc}
\nn
      (n+1) a_n + n (-n^2 +\od n -\od +9) b_n \qquad
\nn	    	        + (n-1) c_n \eql \ldots,
\ear
    where $n \geq 3$ and
\[
     a_n = A_{n+1}/A_2,\qquad b_n = R_{n-1}/R_0, \qquad c_n = f_n/f_1.
\]
    The determinant of \eqs(\ref{abc}) with respect to $a_n,\ b_n,\ c_n$ is
    a multiple (with a nonzero coefficient) of
\beq
       n[ \od (n^2 -2n -7) + n^2-9 ].                     \label{det_abc}
\eeq
    This quantity is nonzero for all values of $\od$ and $n$ of interest:
    it is negative for $n=3$ and positive for $n>3$. Therefore
    \eqs (\ref{abc}) can be consecutively solved for all $n$, leading to a
    $C^\infty$ solution to the field equations (\ref{00q})--(\ref{02q}).
    The proof is completed. $\DAL$
\medskip

    Thus the only kind of STT configurations admitting CC-II is a $D\geq
    4$ Kantowski-Sachs cosmology consisting of two T-regions (in fact,
    epochs, since $\rho$ is a temporal coordinate), separated by a
    second-order horizon. The qualitative behaviour of the metric function
    $\cA(q)$ is shown by the curve 6b in \fig 1, and the Carter-Penrose
    diagram is 6b in \fig 2. We shall see in the next section that this
    conclusion does not change even if the same solution undergoes one more
    CC.

    Unlike CC-I, the present case requires specific properties of the
    potential $U$. It cannot vanish everywhere, but must behave as
    $\const\cdot (\phi-\phi_0)$ near $\phi_0$.

    Moreover, when these requirements are satisfied, the solution that
    realizes a CC-II is even more special than in CC-I. Indeed, in the above
    series expansion, only the constant $f_1$ is arbitrary, whereas $A_2$
    and $R_0$ are expressed via $U_{f0}$, a constant depending on the
    potential in the theory chosen.

\section{Global properties of conformally continued solutions}        

\subsection{Horizon dispositions}

    A solution to the STT equations may {\it a priori\/} undergo a number of
    CCs, so that each region of \MJ\ between adjacent surfaces \Str\ is
    conformally equivalent to some \ME. However, the global properties of
    \MJ\ with CCs turn out to be not so diverse as one might expect. The
    main restriction is that Theorems 2 and 2a on horizon dispositions,
    which have been proved for \ME, actually hold in \MJ.

    A key point for proving this is the observation that the quantity
    $B = A/\rho^2 = \cA/q^2$ is insensitive to
    conformal mappings and is thus common to \MJ\ and \ME\ equivalent to a
    given part of \MJ. We have here presented $B$ in terms of the metrics
    (\ref{dsE}) and (\ref{dsJ}), respectively, so that it
    may be treated as a function of the quasiglobal coordinates
    $\rho$ in \ME\ or $q$ in \MJ.

    A horizon of order $k$ in \MJ\ is evidently a zero of the same order of
    the function $B(q)$, and between zeros (if more than one) there must be
    maxima and minima. Theorem 2 rests on the fact that $B(\rho)$ cannot
    have a regular minimum in \ME\ due to \eq (\ref{02d'}). Hence it follows
    that $B(q)$ cannot have a regular minimum in any region of \MJ\
    equivalent to a particular \ME. A minimum can thus take place only
    on a transition surface \Str\ between two such regions. Consider
    \eq (\ref{02q}) rewritten in terms of $B(q)$ (an analogue of
    (\ref{02d'}) in \MJ):
\bearr   \nq                                                    \label{B_q}
     f\bigl[ R^4 B_{qq} + (\od+2)R^3 R_q B_q + 2(\od-1)\bigr]
     						+ R^4 f_q B_q =0,
\nnn
\ear
    Assuming that $q=0$ is \Str\ and simultaneously an extremum of $B(q)$
    and taking into account Theorems 3 and 4, we can suppose that the Taylor
    expansions of $B$, $f$, $R$ near $q=0$ begin with the terms
\bear
     	B \eql B_0 + \half B_2 q^2 +\ldots,
\nn                                                       \label{ser-q}
	f \eql f_1 q + \ldots,
\nn
	R \eql R_0 + R_1 q + \ldots
\ear
    with nonzero $f_1$ and $R_0$. Substituting (\ref{ser-q}) into
    (\ref{B_q}), we find in the senior order of magnitude $O(q)$:
\beq
	B_2 = - (\od-1)/R_0^4.
                            \label{B_2}
\eeq
    Thus $q=0$ is a maximum of $B(q)$ --- in particular, if $B_0=0$, we are
    dealing with CC-II, and \Str\ separates two T-regions.

    The lack of minima of $B(q)$ means that there can be at most two
    simple zeros, with $B > 0$ between them, or one double zero outside
    which $B < 0$.

    We thus obtain the following theorem, extending Theorem 2 to Jordan
    frames of arbitrary STT:

\Theorem{Theorem 5}
   {In the theory (\ref{act-J}), $D\geq 4$, under the conditions
    $l(\phi) > 0$ and $h(\phi) > 0$, configurations with the metric
    (\ref{dsJ}) and the field $\phi=\phi(q)$ can have at most two simple
    horizons (and there is then an {\rm R}-region between them), or
    one double horizon separating two {\rm T}-regions.  }

    There certainly can be a single simple horizon, as in the \Sch\ or
    \dS\ space-times, or no horizons at all. Theorem 5 means that, precisely
    as in GR, the list of possible causal structures of scalar-vacuum
    configurations in STT is exhausted by the list presented in \sect 3 for
    systems with a cosmological constant in $D$-dimensional GR.

    The situation is still simpler for $D=3$: in this case a CC (more
    precisely, CC-I) is only possible under the condition $B=\const\ne 0$,
    which excludes horizons. A single horizon can exist, but there is then
    no CC.

\subsection{Multiple CCs, singularities and wormholes}

    A full classification of STT solution behaviours is beyond the scope of
    this paper; we will instead outline some new features appearing in STT
    formulated in \MJ\ as compared with GR or with the Einstein-frame
    formulation of STT, in particular, in connection with conformal
    continuations. For simplicity, we will use the STT parametrization such
    that $h\equiv 1$.

    A singularity in \MJ\ can emerge due to the behaviour of the conformal
    factor $F = |f|^{-2/\od}$ at points where \ME\ is regular. The most
    evident case is that $f\to \infty$ at some value of $\phi$, then $F\to
    0$, and we obtain both $\cA = g_{tt}\to 0$ and $R^2 =
    g_{\theta\theta}\to 0$.

    Two kinds of singularities can appear in an ``anti-gravitational''
    region where $f < 0$, that is, beyond a CC surface. The first one occurs
    if $\phi$ blows up while $\psi$ is finite. Using the relation
    (\ref{phi-wag}) between $\phi$ and $\psi$, one easily finds that this is
    only possible when, at large $|\phi|$,
\beq
     f(\phi) \approx  -\frac{D-2}{4(D-1)}\,\phi^2,       \label{as-conf}
\eeq
    i.e., when the $\phi$ field is asymptotically conformally coupled to the
    curvature. The singularity is then again connected with $F\to 0$
    that leads to zero $A(q)$ and $R(q)$.

    Another kind of singularity is more generic and occurs where
\beq
     l(\phi) = f + \frac{D-1}{D-2}\biggl(\frac{df}{d\phi}\biggr)^2  \to 0.
							\label{l_to_0}
\eeq
    Recall that we adhere to the assumption $l > 0$ in the whole paper.
    In the case (\ref{l_to_0}), the conformal factor $F$ is finite
    provided $f\ne 0$ but its derivatives generically blow up:
\[
    \frac{dF}{du} = - \frac{2}{\od\sqrt{l (\phi)}}
    			         \frac{df}{d\phi} \frac{d\psi}{du}
\]
    ($u$ is any admissible coordinate in \ME), which can only be
    finite if $d\psi/du \to 0$ at the same $u$. So, a value of $\phi$ where
    $l\to 0$ is generically a singular sphere of finite radius. Recall also
    \eq(\ref{trace}) showing that, again generically, the scalar curvature
    $\cR$ is infinite where $l=0$. Special solutions where such a sphere is
    regular are not completely excluded but are not considered here.

    Under the assumption $l > 0$, {\it there cannot be more than two values
    of $\phi$ where CCs are possible}, i.e., those where $f=0$ and
    $df/d\phi\ne 0$; if they do exist, one has $f>0$ between them.
    Indeed, if the function $f(\phi) < 0$ between two zeros, it has to pass
    a minimum where $df/d\phi=0$, hence $l <0$, contrary to our assumption.

    This does not mean, however, that an STT solution cannot contain more
    than two CCs. The point is that $\phi$ as a function of the radial
    coordinate is not necessarily monotonic, so there can be two or more CCs
    corresponding to the same value of $\phi$; see the second example in
    \sect 6. But there can be no more than one CC-II due to Theorem 5:
    other CCs in the same solution, if any, belong to type CC-I and occur in
    a T-region.

    Though, multiple CCs can appear in rather special (if not artificial)
    situations since the very existence of a CC imposes restrictions
    on the solution parameters, such as (\ref{CC-g}) for CC-I.  A transition
    surface $\Str \in \MJ$ corresponds to a singularity $r=0$ in \ME,
    therefore an Einstein-frame manifold \ME, describing a region between
    two transitions, should contain ``two centres'', more precisely, two
    values of the radial coordinate (say, $\rho$) at which $r(\rho)=0$. This
    property, resembling that of a closed cosmological model, is quite
    generic due to $r''\leq 0$ in \eq (\ref{01d}), but a special feature is
    that the conditions (\ref{CC-g}) should hold at both centres.

    Another generic (and more usual) behaviour of \ME\ is that $r$ varies
    from zero to infinity. Let there be a family of such static solutions
    and an STT with $f(\phi)$ having a simple zero. Then, by Theorem 3,
    there is a subfamily of solutions admitting CC-I. A particular solution
    from this subfamily can come beyond \Str\ either to one of the
    above-mentioned types of singularities, or, if ``everything is quiet'',
    to another spatial asymptotic and will then describe a static,
    traversable wormhole. Each of the asymptotics can be either flat (if
    $U(\phi)\to 0$), or \adS\ (if $U(\phi) \to\const <0$). Thus wormholes are
    among generic structures that emerge due to conformal continuations.

\section{Examples}

    We will present two explicit examples of configurations with CC-I.
    The first example is well known and is given here to illustrate the
    generic character of wormholes appearing due to CC. The second one is a
    3-dimensional periodic structure with an infinite sequence of CCs.

\subsection {Conformal scalar field in GR}

    Conformal scalar field in GR can be viewed as a special case of STT,
    such that, in \eq (\ref{act-J}), $D=4$ and
\beq                                                       \label{conf}
    f(\phi) = 1 - \phi^2/6, \qquad  h(\phi)=1, \qquad U(\phi) =0.
\eeq
    After the transformation (\ref{g-wag}) $g\mn = F(\psi)\og\mn$ with
\bear
       \phi \eql \sqrt{6} \tanh (\psi+\psi_0)/\sqrt{6}),\cm \psi_0=\const,
\nn							\label{phi-conf}
      F(\psi) \eql \cosh^2[(\psi+\psi_0)/\sqrt{6}],
\ear
    we obtain the action (\ref{act-d}) with $D=4$ and $V\equiv 0$,
    describing a minimally coupled massless scalar field in GR.
    The corresponding \ssph\ solution is well known: it is the Fisher
    solution \cite{fisher}. In terms of the harmonic radial coordinate
    $u\in \R_+$, specified by the condition
    $g_{uu}=-g_{tt}(g_{\theta\theta})^2$, the solution is \cite{br73}
\bear
    ds^2_{\rm E} \eql \e^{-2mu}dt^2                       \label{fish1}
     	         	- \frac{k^2\e^{2mu}}{\sinh^2(ku)}
        \biggl[\frac{k^2 du^2}{\sinh^2(ku)} + d\Omega^2\biggr],
\nn
     \psi \eql C u,
\ear
    where $d\Omega^2 = d\theta^2 + \sin^2 \theta d \varphi^2$,
    $m$ (the mass), $C$ (the scalar charge), $k>0$ and $u_0$ are
    integration constants, and $k$ is expressed in terms of $m$ and $C$:
\beq
      k^2 = m^2 + C^2/2.                               \label{k-fish}
\eeq

    In case $C=0$, $k=m$ we recover the \Sch\ solution, as is easily verified
    using the coordinate $\rho= 2k/(1-\e^{-2ku})$. The metric (\ref{fish1})
    takes the form (\ref{dsE}) with $A(\rho) = (1-2k/\rho)^{m/k}$.

    Another convenient form of the solution is obtained in isotropic
    coordinates: with $y = \tanh (ku/2)$, \eqs (\ref{fish1}) are converted to
\bear
    ds^2_{\rm E} \eql                                    \label{fish2}
		    A(y)\,dt^2
       - \frac{k^2(1-y^2)^2}{y^4 A(y)} (dy^2 + y^2 d\Omega^2),
\nn
     A(y) \eql \biggl|\frac{1-y}{1+y}\biggr|^{2m/k};
\cm
     \psi =\frac{C}{k}\ln \biggl|\frac{1+y}{1-y}\biggr|.
\ear

    The solution is asymptotically flat at $u\to 0$ ($y \to 0$), has
    no horizon when $C\ne 0$ (as should be the case according to the no-hair
    theorem) and is singular at the centre ($u\to \infty$, $y\to 1-0$,
    $\psi\to\infty$).

    A feature of importance is the invariance of (\ref{fish2})
    under the inversion $y \mapsto 1/y$, noticed probably for the first
    time by Mitskevich \cite {mitz}. Due to this invariance, the solution
    (\ref{fish2}) considered in the range $y > 1$ describe quite a
    similar configuration, but now $y\to \infty$ is a flat asymptotic and
    $y\to 1+0$ is a singular centre. An attempt to unify the two ranges of
    $y$ is meaningless due to the singularity at $y = 1$. We shall see that
    such a unification, leading to a wormhole, is achieved when the
    singularity is smoothed out in \MJ\ in case $C=\sqrt{6}m$ due to the
    conformal factor.

    The Jordan-frame solution for (\ref{conf}) is described by the metric
    $ds^2 = F(\psi) ds_{\rm E}^2$ and the $\phi$ field according to
    (\ref{phi-conf}). It is the conformal scalar field solution \cite{bbm70,
    bek74}, its properties are more diverse and can be described as
    follows (putting, for definiteness, $m>0$ and $C>0$):

\medskip\noi
{\bf 1.} $C < \sqrt{6}m$. The metric behaves qualitatively as in the Fisher
    solution:  it is flat at $y\to 0$ ($u\to 0$), and both $g_{tt}$ and
    $r^2=|g_{\theta\theta}|$ vanish at $y\to 1$ ($u\to\infty$) --- a singular
    attracting centre. A difference is that here the scalar field is finite:
    $\phi\to \sqrt{6}$.

\medskip\noi
{\bf 2.} $C > \sqrt{6}m$. Instead of a singular centre, at $y\to 1$
    ($u \to \infty$) one has a repulsive singularity of infinite radius:
    $g_{tt}\to \infty$ and $r^2 \to \infty$. Again $\phi\to \sqrt{6}$.

\medskip\noi
{\bf 3.} $C = \sqrt{6}m$, $k=2m$. Now the metric and $\phi$ are
    regular at $y = 1$; this is \Str, and the coordinate $y$ provides a
    continuation. The solution acquires the form
\bear                    \nq
     ds^2 \eql \frac{(1{+}yy_0)^2}{1-y_0^2}\biggl[\frac{dt^2}{(1{+}y)^2}
               -\frac{m^2(1{+}y)^2}{y^4}(dy^2+y^2 d\Omega^2)\biggr],
\nn
     \phi \eql \sqrt{6} \frac{y+y_0}{1 + yy_0},          \label{con-y}
\ear
    where $y_0 = \tanh (\psi_0/\sqrt{6})$. The range $u\in \R_+$,
    describing the whole manifold \ME\ in the Fisher solution, corresponds
    to the range $0 < y < 1$, describing only a region $\MJ'$ of the
    manifold \MJ of the solution (\ref{con-y}). The properties of the
    latter depend on the sign of $y_0$ \cite{br73}. In all cases, $y=0$
    corresponds to a flat asymptotic, where $\phi \to \sqrt{6}y_0$, $|y_0|
    < 1$.

\medskip\noi
{\bf 3a:} $y_0 < 0$. The solution is defined in the range $0 < y <1/|y_0|$.
    At $y=1/|y_0|$, there is a naked attracting central singularity:
    $g_{tt}\to 0$, $r^2\to 0$, $\phi\to\infty$. Such singularities have
    been mentioned in \sect 5.2 as a characteristic feature of solutions for
    conformally and asymptotically conformally coupled scalar fields, see
    \eq (\ref{as-conf}).

\medskip\noi
{\bf 3b:} $y_0 > 0$. The solution is defined in the range $y\in \R_+$.
    At $y\to\infty$, we find another flat spatial infinity, where
    $\phi\to \sqrt{6}/y_0$, $r^2\to\infty$ and $g_{tt}$ tends to a finite
    limit.  This is a {\sl wormhole solution\/}, found for the first time
    in \Ref{br73} and recently discussed by Barcelo and Visser \cite{viss}.

\medskip\noi
{\bf 3c:} $y_0=0$, $\phi=\sqrt{6}y$, $y\in \R_+$. In this case it is helpful
    to pass to the conventional coordinate $r$, substituting $y=m/(r-m)$.
    The solution
\bear
     ds^2 \eql (1-m/r)^2{dt^2} - \frac{dr^2}{(1-m/r)^2} -r^2d\Omega^2,
\nn
     \phi \eql \sqrt{6}m/(r-m)                              \label{con-bh}
\ear
    is the well-known BH with a conformal scalar field \cite{bbm70,bek74}.
    The infinite value of $\phi$ at the horizon $r=m$ does not make the
    metric singular since, as is easily verified, the energy-momentum tensor
    remains finite there. This solution turns out to be unstable under
    radial perturbations \cite{bk78}.

    Case 3 belongs to variant III in the classification of \sect
    3.2, and the whole manifold \MJ\ can be represented as the union
\beq
	\MJ = \MJ' \cup \Str \cup \MJ''
\eeq
    where $\MJ'$ is the region $y<1$, which is, according to
    (\ref{phi-conf}), in one-to-one correspondence with the manifold \ME\
    of the Fisher solution (\ref{fish1}).  The
    ``antigravitational'' ($f(\phi) < 0$) region $\MJ''$
    ($y > 1$) is in similar correspondence with another ``copy'' of the
    Fisher solution, where, instead of (\ref{phi-conf}),
\beq
        \phi = \sqrt{6} \coth (\psi/\sqrt{6}),  \quad     \label{phi-con'}
      F(\psi)=\sinh^2(\psi/\sqrt{6}).
\eeq

    Static wormhole solutions have also been found \cite{viss} for
    more general nonminimally coupled massless scalar fields in GR,
    represented as STT where
\beq                                                       \label{conf-}
     f(\phi) = 1 - \xi\phi^2, \qquad  h(\phi)=1, \qquad U(\phi) =0
\eeq
    with $\xi=\const > 0$. In full agreement with the observations of \sect
    5.2, there appear wormholes similar to (\ref{con-y}), but in
    case $\xi < 1/6$ some of the conformally continued solutions possess
    singularities connected with $l(\phi) = 1 - \xi(1-6\xi) \phi^3 \to 0$.
    In case $\xi > 1/6$ all continued solutions describe wormholes.

    All the wormhole solutions mentioned in this section prove to be
    unstable under radial perturbations \cite{bg01}, which seems to be
    a common feature of transitions to regions with $f < 0$, where the
    effective gravitational constant becomes negative. This problem deserves
    further study. A similar instability was pointed out by Starobinsky
    \cite{star81} for cosmological models with conformally coupled scalar
    fields.

\subsection{A solution with multiple CCs}

    Trying to obtain a simple analytical solution, we choose $D=3$ and the
    functions $f(\phi)$ and $h(\phi)$ in the action (\ref{act-J})
    corresponding to conformal coupling:
\beq
     f (\phi) = 1-\phi^2/8, \cm h (\phi) \equiv 1.           \label{conf-3}
\eeq
    The $\phi-\psi$ connection according to (\ref{phi-wag})
    (with a proper choice of the integration constant) and the
    conformal factor $F(\psi)$ in (\ref{g-wag}) may be written as
\bear
    \phi \eql \sqrt{8}\tanh (\psi/\sqrt{8}),                \label{phi+}
\nn
    F(\psi) \eql \cosh^2 (\psi/\sqrt{8}) \cm {\rm for}\quad \phi^2 < 8,
\yy
    \phi \eql \sqrt{8}\coth (\psi/\sqrt{8}),                \label{phi-}
\nn
    F(\psi) \eql \sinh^2 (\psi/\sqrt{8}) \cm {\rm for}\quad \phi^2 > 8.
\ear
    A solution in \MJ, including regions with $|\phi|$ larger and smaller
    than $\sqrt{8}$, is built from solutions in different \ME\ with
    CC through the surfaces \Str\ on which $\phi^2=8$.

    Let us first construct a solution in the Einstein frame, to be put into
    correspondence any of the regions (\ref{phi+}) and (\ref{phi-}) in \MJ,
    in such a way as to avoid $\psi=0$, since otherwise we shall
    encounter a singularity due to $F(\psi)=0$ in the region (\ref{phi-}).
    We will use the metric in the form (\ref{dsE}) and \eqs
    (\ref{00d})--(\ref{02d}) with $\od=1$.

    As follows from the aforesaid, to provide a CC we must choose a
    solution to (\ref{02d}) in the form
\beq
	   A(\rho) = c_A\, r^2(\rho), \cm c_A = \const,          \label{A-3}
\eeq
    where $c_A>0$ and $c_A<0$ corresponds to static and cosmological
    solutions, respectively.

    By (\ref{r_trans}), near \Str\ ($\rho=\rho_c$) the function $r(\rho)$
    behaves as $(\rho-\rho_c)^{1/D}$.  Accordingly, let us choose this
    function as follows:
\bearr
       r(\rho) = r_0 (1-x^4)^{1/3}, \qquad x=\rho/\rho_0,
   \nnn
         		\qquad  \rho_0 = \const >0.           \label{r-3}
\ear
    Thus a CC-I can occur at $x=\pm 1$.

    Now, \eq (\ref{01d}) makes it
    possible to find $\psi(\rho)$, or equivalently $\psi(x)$:
\beq
      \pm 3\frac{d\psi}{dx} = \frac{2x\sqrt{9-x^4}}{1-x^4}.  \label{psi_x}
\eeq
    Choosing the plus sign and integrating, we find
\beq
     \psi(x) = -\frac{\sqrt{2}}{3}\ln (1-x^2)
     					+ \psi_1(x) + \psi_0  \label{psi3}
\eeq
    where $\psi_0$ is an integration constant while the function $\psi_1$
    is analytic and finite for all $|x|\leq 1$:
\bearr
     \psi_1(x) = \arcsin (x^2/3)                             \label{psi_1}
\nnn\cm
	   + \sqrt{2}\,\ln
	   	   \frac{(1+x^2)\bigl[9-x^2+\sqrt{8(9-x^4)}\bigr]}
	   		{9+ x^2 + \sqrt{8(9-x^4)}}.
\ear
    The potential $V(\psi)$ as a function of $x$ is found from (\ref{00d}):
\beq
      V(\psi) = 2 \frac{c_Ar_0^2}{\rho_0^2}                   \label{V-3}
      				\frac{x^2}{(1-x^4)^{1/3}}.
\eeq
    This completes the solution in \ME, specified in the range
    $-1<x<1$. All the constituent functions are even. The potential
    $V(\psi)$ is well-defined due to monotonicity of $\psi(x)$ in each
    half-range $0 < |x| <1$. At possible CCs, $x=\pm 1$, $\psi \to \infty$,
    and the minimum value of $\psi$ is $\psi(0) = \psi_0$; assuming
    $\psi_0 > 0$, we make sure that $\psi > 0$ everywhere.

    The corresponding solutions in the Jordan frame are different for
    $\phi < \sqrt{8}$ and $\phi > \sqrt{8}$, according to (\ref{phi+}) and
    (\ref{phi-}) where we put for certainty $\phi > 0$.

\Picture{85}
{\unitlength=.55mm
\special{em:linewidth 0.4pt}
\linethickness{0.4pt}
\begin{picture}(152.00,150.00)(5,4)
\put(9.00,20.00){\line(1,0){143.00}}
\put(20.00,20.00){\line(0,1){130.00}}
\bezier{144}(100.00,87.00)(110.00,102.00)(120.00,87.00)
\bezier{108}(13.00,27.00)(17.00,15.00)(28.00,24.00)
\bezier{72}(28.00,24.00)(35.00,29.00)(43.00,24.00)
\bezier{116}(43.00,24.00)(54.00,16.00)(67.00,24.00)
\bezier{72}(82.00,24.00)(74.00,29.00)(67.00,24.00)
\bezier{104}(82.00,24.00)(92.00,15.00)(97.00,27.00)
\bezier{96}(125.00,24.00)(127.00,16.00)(141.00,24.00)
\bezier{156}(20.00,125.00)(28.00,125.00)(35.00,95.00)
\bezier{176}(35.00,95.00)(44.00,63.00)(55.00,64.00)
\bezier{204}(75.00,103.00)(66.00,64.00)(55.00,64.00)
\bezier{128}(75.00,103.00)(81.00,125.00)(90.00,125.00)
\bezier{160}(90.00,125.00)(96.00,125.00)(100.00,91.00)
\bezier{164}(130.00,125.00)(123.00,125.00)(120.00,91.00)
\bezier{668}(100.00,91.00)(110.00,8.00)(120.00,91.00)
\bezier{100}(130.00,125.00)(137.00,125.00)(144.00,108.00)
\bezier{176}(20.00,125.00)(14.00,125.00)(10.00,87.00)
\bezier{376}(20.00,76.00)(55.00,45.00)(90.00,76.00)
\bezier{60}(90.00,76.00)(96.00,81.00)(100.00,87.00)
\bezier{132}(120.00,87.00)(131.00,73.00)(144.00,66.00)
\bezier{48}(20.00,76.00)(12.00,84.00)(12.00,85.00)
\bezier{108}(125.00,24.00)(121.00,37.00)(115.00,25.00)
\bezier{88}(97.00,27.00)(101.00,36.00)(105.00,25.00)
\bezier{88}(105.00,25.00)(110.00,15.00)(115.00,25.00)
\bezier{60}(13.00,27.00)(9.00,34.00)(6.00,28.00)
\bezier{112}(10.00,87.00)(7.00,67.00)(5.00,60.00)
\linethickness{0.18pt}
\put(90.00,150.00){\line(0,-1){130.00}}
\put(130.00,150.00){\line(0,-1){130.00}}
\put(19.00,40.00){\line(1,0){2.00}}
\put(19.00,60.00){\line(1,0){2.00}}
\put(21.00,80.00){\line(-1,0){2.00}}
\put(19.00,100.00){\line(1,0){2.00}}
\put(21.00,120.00){\line(-1,0){2.00}}
\put(19.00,140.00){\line(1,0){2.00}}
\put(18.00,16.00){\makebox(0,0)[cc]{$q_0$}}
\put(89.00,16.00){\makebox(0,0)[cc]{$q_1$}}
\put(129.00,16.00){\makebox(0,0)[cc]{$q_2$}}
\put(48.00,11.00){\makebox(0,0)[cc]{${\mathbb M}_0$}}
\put(108.00,11.00){\makebox(0,0)[cc]{${\mathbb M}_1$}}
\put(147.00,11.00){\makebox(0,0)[cc]{${\mathbb M}_2$}}
\put(15.00,40.00){\makebox(0,0)[cc]{1}}
\put(15.00,60.00){\makebox(0,0)[cc]{2}}
\put(15.00,100.00){\makebox(0,0)[cc]{4}}
\put(13.00,120.00){\makebox(0,0)[cc]{5}}
\put(15.00,140.00){\makebox(0,0)[cc]{6}}
\put(46.00,107.00){\makebox(0,0)[cc]{$R(q)/r_0$}}
\put(141.00,61.00){\makebox(0,0)[cc]{$\phi(q)$}}
\put(63.00,30.00){\makebox(0,0)[cc]{$U\bigl(\phi(q)\bigr)$}}
\put(14.00,72.00){\makebox(0,0)[cc]{$\sqrt{8}$}}
\multiput(15.00,76.00)(6,0){22}{\line(1,0){3.00}}
\end{picture}
}
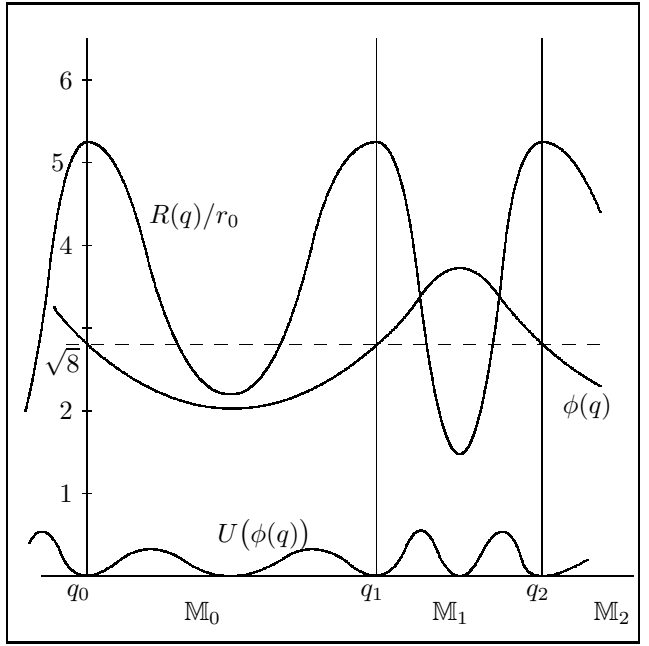
{The behaviour of $\phi(q)$, $R(q)$, $U(\phi(q))$ in the model
 of \sect 6.2 with multiple CCs, in case $\psi_0 = \sqrt{8}$, on a
 segment of the infinite sequence of regions $\M_i$. The latter are
 separated by vertical lines $q=q_i$ corresponding to the surfaces \Str..
 The potential $U(\phi)$ is shown in an arbitrary scale.  }

    The solution in \MJ\ for $\phi < \sqrt{8}$,  obtained from \ME\ using
    (\ref{phi+}), occupies in \MJ\ a certain region $\M_0$ parametrized by
    $x\in (-1.+1)$. The solution can be continued through the surface,
    say, $x=1$ to $\phi > \sqrt{8}$; to do that one can, i.e., consider the
    metric coefficients as functions of $\phi$ and analytically continue
    them beyond the value $\phi=\sqrt{8}$. However, one cannot do that
    explicitly since the transcendental equation (\ref{psi3}) cannot be
    resolved with respect to $x$.

    In the new region $\M_1$, which can again be parametrized by $x \in
    (-1, +1)$, another ``copy'' of the Einstein-frame solution
    (\ref{A-3})--(\ref{V-3}) is valid. To make sure that this is the same
    solution as the one used in $\M_0$, let us consider the transition
    between them and recall the proof of Theorem 3 (sufficiency). Namely,
    there is a unique solution in \MJ\ near \Str\ (in the form of a power
    series in $q$) if the functions $f(\phi)$ and $U(\phi)$ and the
    constants $A_0$, $R_0$ and $f_1$ are specified. In our case $f(\phi)$
    is given, $U(\phi)$ is not known explicitly but its existence follows
    from the existence of the analytic continuation in terms of $\phi$. So
    it is sufficient to show that the constants $A_0$, $R_0$ and $f_1$,
    calculated as the limiting values of $\cA = AF$, $R= r\sqrt{F}$ and
    $df/dq$, respectively, from the two solutions, are finite and coincide
    with each other. A direct inspection shows that this is indeed the case
    if all the parameters of the solutions are the same, including
    $\psi_0$.

    In $\M_1$ the field $\phi$ reaches its maximum at $x=0$,
\beq
    \phi_{\max} = \sqrt{8} \coth (\psi_0/\sqrt{8}),        \label{phimax}
\eeq
    and returns to the value $\sqrt{8}$ on the other end of the region. One
    more CC leads to one more region $M_2$ with $\phi < \sqrt{8}$ and so
    on. The same happens starting from $x=-1$ of the initial region $\M_0$.
    We obtain {\it an infinite sequence of regions\/} $M_i$, $i\in \Z$,
    where adjacent regions are connected by CCs. In regions with even and
    odd numbers $i$, one has $\phi < \sqrt{8}$ and $\phi > \sqrt{8}$,
    respectively. Each region $\M_i\in \MJ$ corresponds to its own
    Einstein-frame manifold $\ME_i$, described by the solution
    (\ref{A-3})--(\ref{V-3}) with singularities at $x\to\pm 1$.

    The whole manifold \MJ\ can be parametrized by a unique ``radial''
    coordinate: it can be, e.g., the quasiglobal coordinate $q$ used in
    Theorem 3, or the proper length $\ell = \int dq/\sqrt{\cA}$; both
    quantities take finite values on the transition surfaces from $\M_i$ to
    $\M_{i+1}$ and change monotonically inside $\M_i$.

    The manifold \MJ\ is thus nonsingular and has the topology
    $\R \times \R \times \S^1$: an infinitely long static tube with a
    periodically changing diameter. This is true if $c_A >0$, when we
    deal with a static model. One can identify any two $\M_i$
    with the numbers $i$ of equal parity, and this leads to the
    topology $\R \times \S^1 \times \S^1$, in other words, 2-torus times
    the time axis. If $c_A < 0$, then $q$ is a temporal coordinate,
    $\ell$ becomes proper time, and the solution describes a
    (2+1)-dimensional cosmology with a periodically and isotropically
    (since $A \propto R^2$) changing scale factor. The spatial section is
    $\R\times \S^1$, but any two points on the $t$ axis ($t$ is now
    spacelike) may be identified, and we obtain a periodically
    ``breathing'' 2-torus.

    The properties of the model are illustrated in Fig.\,3.

\subsection*{Acknowledgements}

I am grateful to Vitaly Melnikov and Georgy Shikin for helpful discussions.
The work was supported in part by the Russian Foundation for Basic
Research and the Russian Ministry of Industry, Science and Technologies.

\end{document}